\documentclass[aps,pre,preprint,showpacs,superscriptaddress]{revtex4}

\newcommand{\be}{\begin{equation}}
\newcommand{\ee}{\end{equation}}
\newcommand{\bea}{\begin{eqnarray}}
\newcommand{\eea}{\end{eqnarray}}
\renewcommand{\d}{\text{d}}

\usepackage{amsmath,bbm,mathrsfs,graphics}
\DeclareMathAlphabet{\mathpzc}{OT1}{pzc}{m}{it}

\begin{document}

\title{Poisson-to-Wigner crossover transition in the
  nearest-neighbor spacing statistics of random points on fractals}

\author{Jamal Sakhr}
\altaffiliation[Present address: ]{Department of Physics, Harvard University, 
Cambridge, Massachusetts 02138 USA} 
\affiliation{Department of Physics and Astronomy, McMaster University,
Hamilton, Ontario, Canada L8S~4M1}
\author{John M. Nieminen}
\affiliation{NDI (Northern Digital Inc.), 103 Randall Drive, 
Waterloo, Ontario, Canada N2V~1C5}

\date{\today}

\begin{abstract}
We show that the nearest-neighbor spacing distribution for a model
that consists of random points uniformly distributed on a self-similar 
fractal is the Brody distribution of random matrix  
theory. In the usual context of Hamiltonian systems, the Brody parameter 
does not have a definite physical meaning, but in the model considered here, 
the Brody parameter is actually the fractal dimension. Exploiting this
result, we introduce a new model for a crossover transition 
between Poisson and Wigner statistics: random points on a continuous family of
self-similar curves with fractal dimension between 1 and 2. The implications 
to quantum chaos are discussed, and a connection to conservative classical 
chaos is introduced.  
\end{abstract}


\maketitle

It is well known that the spectral statistics of almost all 
quantum systems, whose classical analogs are 
chaotic, are described quantitatively by 
Gaussian ensembles of random matrices \cite{Bohigas,Stockmann,Haake,Reichl}. 
Perhaps less well known is the fact that the eigenvalue spacing
distribution for the Gaussian orthogonal ensemble of $2\times2$ random 
matrices also describes homogeneous Poisson point 
processes in $\mathbbm{R}^2$ \cite{Haake}. The nearest-neighbor spacing
distribution (NNSD) of random points uniformly distributed on a
line is given by  
\begin{equation} \label{Eqn:Poisson}
    P_{P}(S) = \exp(-S),
\end{equation}
and the NNSD of random points uniformly distributed on a plane
is given by \cite{Haake} 
\begin{equation} \label{Eqn:Wigner}
    P_{W}(S) = \frac{\pi}{2} S \exp\left( -\frac{\pi}{4} S^2 \right), 
\end{equation} 
which is, in fact, the Wigner distribution of random matrix theory (RMT)
 \cite{Mehta}. (In 
Eqs.~(\ref{Eqn:Poisson}) and (\ref{Eqn:Wigner}), $S$ is a  
dimensionless scaled spacing.) 
Evidently, random points on a line are uncorrelated, whereas 
random points on a plane are actually  
correlated (in the sense that the points tend to avoid each other). 
Even so, the latter result is reasonable  
since there is an additional degree of freedom that allows the
points to spread out. Based on this intuitive interdependence between 
``point repulsion'' and dimensionality, and in strict analogy to
energy-level statistics, it is tempting to
indiscriminately conjecture that the nearest-neighbor statistics of 
random points on a fractal set with noninteger dimension 
between $1$ and $2$ are described by an intermediate distribution in-between 
Poisson and Wigner. In fact, this conjecture turns out to be
correct as we shall demonstrate below.

Our interest in the nearest-neighbor statistics of random points on
fractals was sparked by the provocative results (\ref{Eqn:Poisson})
and (\ref{Eqn:Wigner}), and the prospect that random points on fractals 
could be conceptualized as new models of intermediate statistics (between
Poisson and Wigner). We however are not the first authors to consider
the statistical properties of fractal sets. 
Two decades ago, Badii and Politi \cite{BadandPol} studied (for completely
different reasons) the nearest-neighbor distance distribution of
random points on a strange attractor. These authors also 
obtained the probability distribution of
nearest-neighbor distances $\delta$ among $N$ points chosen randomly and  
uniformly on a Cantor set with capacity dimension
$D_0$. Interestingly, they noted (quite tersely) that the
asymptotic distribution (appropriate for large $N$) \cite{BadandPol}
\be\label{NNSDBP}
P_{BP}(\delta,N)=2D_0N(2\delta)^{D_0-1}\exp\left[-N(2\delta)^{D_0}\right], 
\ee
could be ``recognized as a Brody distribution''. Clearly, 
Eq.~(\ref{NNSDBP}) is not the Brody distribution \cite{Brody}, but we 
show that when the nearest-neighbor distance is rescaled by the
mean nearest-neighbor distance (which is a contrivance
familiar to practitioners of RMT) the result is indeed \emph{the} Brody  
distribution. More generally speaking, we show that the Brody distribution 
is the NNSD for points selected uniformly at
random from a self-similar set $K \subset\mathbbm{R}^d$ with
similarity dimension $d_s \ge 1$, and that \emph{the Brody parameter is the
(relative) similarity dimension of $K$} (i.e.~$d_s-1$). 
(In this paper, $d$ is the Euclidean space dimension.) 
The goals of this paper are to derive this result, to introduce a new model 
for a Poisson-to-Wigner crossover transition based on this result, and to 
explicate the physical implications. The derivation
is simple and direct; the result itself is far more interesting than
its proof. A discussion of the physical implications is deferred to the  
conclusion. 

We begin first with the derivation of the spacing distribution. Suppose that
$N$ points of a self-similar set $K \subset \mathbbm{R}^d$ are chosen 
randomly and uniformly. 
The probability $P(s)\d s$ of finding the nearest neighbor to a 
given point at a distance between $s$ and $s+\d s$ is equal to the 
probability of finding \emph{one} of the $(N-1)$ points at a distance
between $s$ 
and $s+\d s$ from the given point \emph{and} the $(N-2)$ remaining points 
at a distance greater than $s$. Let $\mathpzc{P}(s)$ denote 
the probability of finding a point within a distance $s$ of a given point. 
The probability of finding one point at a distance greater than $s$ is then 
$(1-\mathpzc{P}(s))$, and for $(N-2)$ points, the probabilities are 
multiplicative due to the fact that all points are chosen
independently. Thus,  
\be \label{Eqn:NNdistraw}
P(s)\d s = \eta\left(1-\mathpzc{P}(s)\right)^{(N-2)}\d\mathpzc{P}(s),
\ee
where the prefactor $\eta=(N-1)$ accounts for the fact
that the nearest neighbor could be any one of the $(N-1)$ points 
\footnote{Equation (\ref{Eqn:NNdistraw}) \emph{without} the prefactor
    $\eta$ implicitly assumes one of the points (say
    $\mathbf{x}_2$) is the nearest neighbor to the ``given point'' (say 
$\mathbf{x_1}$).  So, \emph{without} $\eta$, we have only
    computed the probability that
    $\mathbf{x}_2$  is at a distance $s$ (the closest distance to
$\mathbf{x}_1$). 
We must now sum all the \emph{equivalent} probabilities that           
the nearest neighbor is at distance $s$ from $\mathbf{x}_1$, with the
recognition that the nearest neighbor in question can be any of the
remaining $(N-1)$ points. Hence, the prefactor $(N-1)$.}, and 
$\d \mathpzc{P}(s)= \mathpzc{P}'(s)\d s$ 
is the probability of finding a point in a shell with inner and outer 
radii $s$ and $s+\d s$ centered about the given point. 
The probability of finding multiple nearest neighbors is ignored since 
the probability of such an event is higher order in $\d s$ 
and is therefore insignificant compared to the probability of finding a 
single nearest neighbor. 

It now remains to specify the probability $\mathpzc{P}(s)$.
Clearly, the typical number of neighbors of a
given point will vary more rapidly with distance from that point as
the dimension increases. The probability $\mathpzc{P}(s)$ is,  
by definition, the ratio of the number of points within some
prescribed distance to the total number of points, that is, 
$\mathpzc{P}(s)=N(s)/N(\mathcal{R})$, where $\mathcal{R}$ is the radius 
of the $d$-dimensional ball that contains all $N$ points. The number
function $N(s)$ can be directly obtained from the so-called 
``mass-radius scaling law'' for fractals 
(see page 40 of Mandelbrot's book \cite{ManBrot}):
$M(r)=M(\mathcal{R}) (r/\mathcal{R})^{D_m}=[M(\mathcal{R})/
  \mathcal{R}^{D_m}]r^{D_m}$. In this formula, 
$M(r)$ and $M(\mathcal{R})$ are the masses contained within
balls of radii $r$ and $\mathcal{R}$,  
respectively, and $D_m$ is the mass dimension. For regular
(i.e.~\emph{strictly} self-similar) fractals, $D_m=d_s$. 
It might seem peculiar to speak of masses here, but it is equivalent
to the concept of numbers of points within balls of a specified radius if
an individual sample point becomes the unit of mass. We can thus equitably
think of the above mass law as a ``number-radius scaling law''. So, the  
probability of finding a point within a 
distance $s$ of a given point is governed by the power law 
\be \label{postulat}
\mathpzc{P}(s)=As^{d_s}, 
\ee
where the coefficient $A=1/\mathcal{R}^{d_s}$, and $d_s \ge 1$ is 
the similarity dimension of $K$. 
(Note that the probability $\mathpzc{P}(s)$ is a unitless 
number since $A$ has units of $1/(\text{length})^{d_s}$.) 
Therefore, Eq.~(\ref{Eqn:NNdistraw}) becomes 
\be \label{Eqn:NNdist}
P(s)\d s = \eta\left(1-As^{d_s}\right)^{(N-2)}Ad_ss^{d_s-1}\d s.
\ee 
Recall that $\d \mathpzc{P}(s)= \mathpzc{P}'(s)\d s=Ad_ss^{d_s-1}\d s$.   
It is straightforward to verify that the probability density $P(s)$ is
already normalized (i.e.~$\int_0^{\mathcal{R}}P(s)\d s=1$). We could
now consider the large $N$ limit of Eq.~(\ref{Eqn:NNdist}), and in
doing so, we can invoke the so-called Poisson approximation
\footnote{If $N$ is large and $N\gg j$, then $(1-x)^{N-j} \approx
  e^{-Nx}$.}, 
and this gives the asymptotic probability density 
\be\label{NNdistlargeN} 
P(s)=NAd_ss^{d_s-1}\exp\left(-NAs^{d_s}\right) \quad     
\text{as}~N\rightarrow\infty. 
\ee
Equation (\ref{NNdistlargeN}) is essentially the distribution obtained
by Badii and Politi in 1985 (c.f.~Eq.~(\ref{NNSDBP}) and note that
$D_0$ in their formula is equivalent to $d_s$ in Eq.~(\ref{NNdistlargeN})). 

Next, we calculate the mean spacing $\bar{s} =\int_0^{\mathcal{R}}
sP(s)\d s$:
\bea\label{Eqn:meanspacing}
\bar{s}&=&{(N-1)d_s\over \mathcal{R}^{d_s}} 
\int_0^{\mathcal{R}}s^{d_s}\left(1-\left({s\over \mathcal{R}}\right)^{d_s}
\right)^{(N-2)}\d s \nonumber \\
&=&\mathcal{R}(N-1)d_s\int_0^1 u^{d_s}
\left(1-u^{d_s}\right)^{(N-2)}\d u \nonumber \\
&=&\mathcal{R}(N-1)\int_0^1v^{1/d_s}(1-v)^{(N-2)}\d v \nonumber \\
&=&\mathcal{R} (N-1) B\left(1+{1/d_s}, N-1\right) \nonumber \\ 
&=&\mathcal{R}\Gamma(N)\Gamma\left(1+{1/d_s}\right)/\Gamma\left(N+{1/d_s} 
\right). \nonumber 
\eea
In the second line, we have made a change of variables 
$u=s/\mathcal{R}$, and in the third line, we have made one further
change of variables $v=u^{d_s}$. The integral in the third line we
recognize as the Beta function $B(\mu,\nu)$ with parameters
$\mu=1+1/d_s$ and $\nu=N-1$, and this then gives the fourth line. We then 
used the usual relation between the Gamma and Beta functions to arrive
at the fifth line. It can be shown that the term 
\be \label{gammaratio}
{\Gamma(N)\over\Gamma\left(N+{1/d_s}\right)}={1\over
  N^{1/d_s}}\left(1+O\left({1\over N}\right)\right) \quad     
\text{as}~N\rightarrow\infty, 
\ee 
and therefore, the asymptotic mean nearest-neighbor spacing 
is, to leading order, 
\be \label{sbar}
\bar{s}={\mathcal{R}\over N^{1/d_s}}~ \Gamma\left({d_s+1\over
  d_s}\right) \quad \text{as}~N\rightarrow\infty.
\ee
Introducing the rescaled spacing $S={s/\bar{s}}$ and 
taking the limit $N\rightarrow\infty$, the distribution $P(s)$ in 
Eq.~(\ref{NNdistlargeN}) becomes the distribution 
\begin{subequations} \label{Eqn:general} 
\be  
  P_B(S;q=d_s-1) =  \alpha d_s S^{d_s-1} \exp\left( -\alpha S^{d_s} \right), 
\label{Eqn:gen1}
\ee 
where
\be 
   \alpha =  \left[ \Gamma \left( \frac{d_s+1}{d_s} \right) \right]^{d_s}
\label{Eqn:alpha_d}.
\ee
\end{subequations} 
(Notice that the rescaling of $s$ by $\bar{s}$ was doubly beneficial;
both explicit dependences on $\mathcal{R}$ and $N$ in Eq.~(\ref{NNdistlargeN})
have been removed.) 
The distribution $P_B(S;q)$ [Eq.~(\ref{Eqn:general})] is, in fact, the
Brody distribution \cite{Brody} with Brody parameter $q$ equal to $d_s-1$ 
\footnote{The Brody distribution is usually written as 
$P_{B}(S;q) = \lambda (q+1) S^{q} \exp\left( -\lambda S^{q+1}
  \right)$, where $\lambda = \left[ \Gamma \left( \frac{q+2}{q+1}
    \right) \right]^{q+1}$ and $q$ is the 
Brody parameter. This reduces to the Poisson distribution 
[Eq.~(\ref{Eqn:Poisson})] when $q=0$ and the 
Wigner distribution [Eq.~(\ref{Eqn:Wigner})] when $q=1$.}. We refer to
the number $d_s-1$ as the \emph{relative} similarity dimension of $K$
since this number is the difference between the similarity dimension
of $K$ and the similarity dimension of a line (the simplest self-similar 
object) which is equal to $1$. 
Equation (\ref{Eqn:general}) is valid for random points on any self-similar 
subset of $\mathbbm{R}^d$ with similarity dimension $d_s \ge 1$. 
If $K$ is a \emph{classical} (nonfractal) self-similar set 
(i.e.~a $d$-dimensional cube), then $d_s=d$ and
Eq.~(\ref{Eqn:general}) reduces to 
the NNSD for a homogeneous Poisson point process in $\mathbbm{R}^d$ 
(see Ref.~\cite{Haake}). 
 
\begin{figure} 
\scalebox{0.653}{\includegraphics*{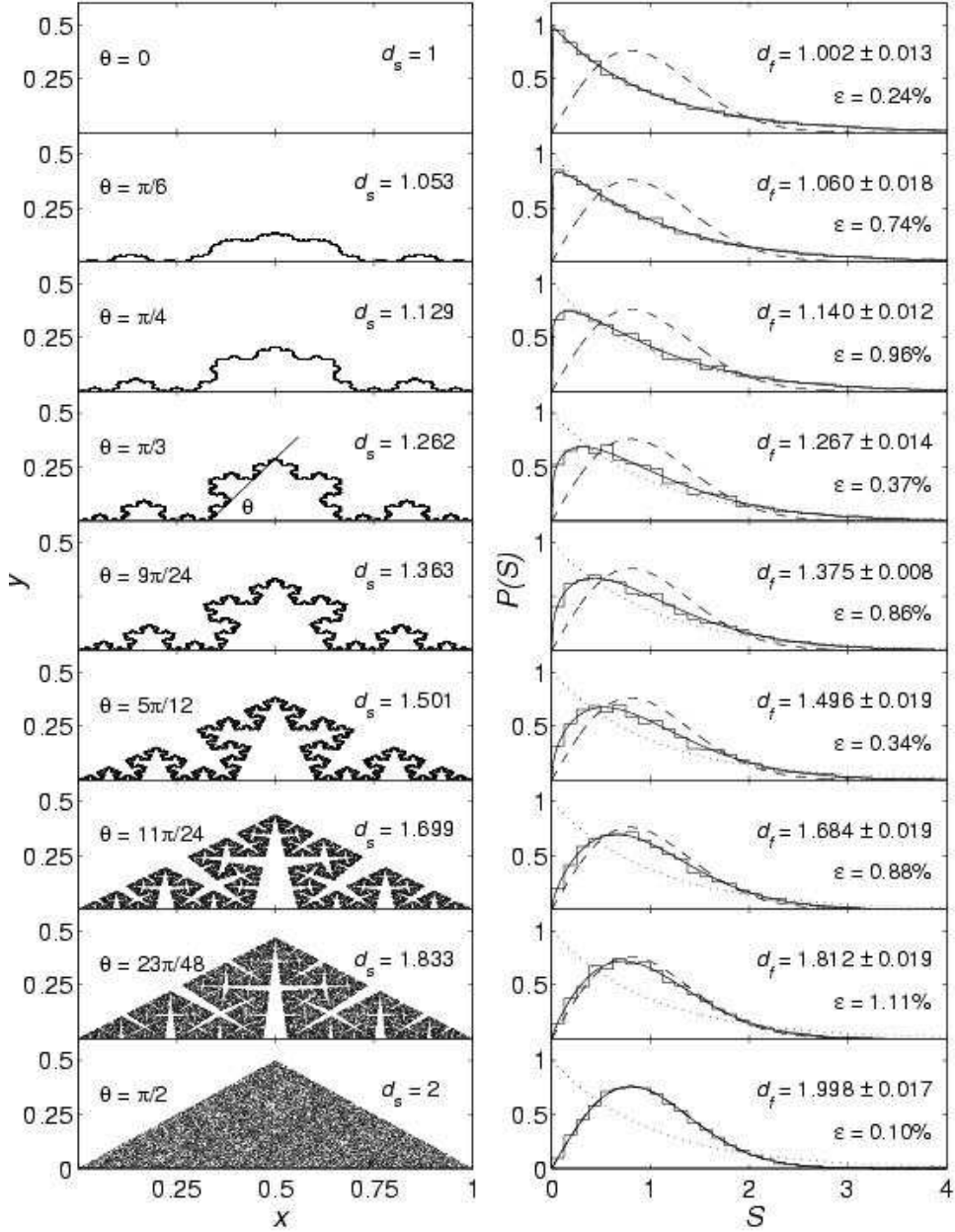}}
\caption{\label{allKoch} A crossover transition between Poisson and 
Wigner statistics resulting from point processes on the family of Koch
fractals  
in $\mathbbm{R}^2$. The left panel shows random points on several fractals  
that belong to the family (each one specified by a particular choice
of the rotation angle $\theta$). 
The exact similarity dimension $d_s$ of each of the fractals (given to
three decimal places) is indicated on each window. 
The right panel shows the corresponding  
NNSD of the points on each fractal. The numerical data 
(i.e.~$d_f=\bar{d}_f\pm\sigma$  
and $\varepsilon$) indicated on each window are as described in the text. 
The dotted, dashed, and solid curves 
are the distributions $P_P(S)$, $P_W(S)$, and $P_B(S;\bar{d}_f-1)$,
respectively.}
\end{figure}

In studies of quantum chaos, the Brody distribution has sometimes been
used (as a purely phenomenological distribution) to describe the 
nearest-neighbor energy-level statistics of quantum systems that undergo 
a direct transition from Poisson-like to Wigner-like statistics as a system  
parameter is varied. (A classic example is the diamagnetic Kepler
system \cite{classi}.) In the present context, a Poisson-to-Wigner transition 
can be realized by considering point processes on a family of self-similar  
sets whose dimension ranges between $1$ and $2$ as some set parameter is  
varied. In actual fact, we are introducing a new model (that does not
involve random matrices) for a Poisson-to-Wigner crossover transition, and 
this model is special since the intermediate statistics are described exactly 
by the Brody distribution. 
As a concrete example, we now study point processes on 
the family of Koch fractals in $\mathbbm{R}^2$.  
These fractals can be thought of as the attractors of
a one-parameter family of iterated function systems (IFSs). 
The similarity transformations defining the IFS involve a rotation
which is conveniently parametrized by the angle $\theta$. When 
$\theta=0$, the attractor is a line, and when $\theta=\pi/2$, the
attractor is the famous Sierpinski-Knopp plane-filling curve, whose
image is a solid isosceles triangle in $\mathbbm{R}^2$ \cite{Sagan}. For
intermediate values [i.e.~$\theta\in(0,\pi/2)$], the attractors are
various self-similar curves of prescribed dimension $d_s \in (1,2)$. 
The nearest-neighbor statistics of the random
points undergo a continuous transition from Poisson to Wigner statistics
(see Fig.~\ref{allKoch}) 
as the self-similar set continuously deforms from a line to a
plane-filling curve (i.e.~as the rotation angle $\theta$ varies
between $0$ and $\pi/2$). 

For clarity, we mention some pertinent numerical details. 
Random points on these fractals were selected using the random
iteration algorithm (RIA) \cite{Barnbook}. 
The distance between a given point $\mathbf{x}_i$ and its nearest neighbor  
is defined (using the Euclidean metric) by  
$s_{i} = \text{min} \left\{\sqrt{(x_i-x_j)^2+(y_i-y_j)^2}\right\}$ 
for $i,j=1,\ldots,N$ ($j \neq i$). 
Although $\{s_i\}$ define a set of spacings, 
the NNSD is actually defined in terms of the scaled spacings
$S_i = s_i / \bar{s}$, 
where $\bar{s}={1\over N}\sum_{i=1}^N s_i$, is the (numerically-calculated) 
mean nearest-neighbor spacing. We constructed the histograms 
by binning all values of $S_i$ and then normalizing the area under the
histogram to unity. 
Each histogram was constructed from one sample of $N=20~000$ random points. 
We mention here that by using all of the points selected by the RIA
in the statistical analysis, we introduce some error due to finite-size 
or edge effects. These errors are statistically insignificant
as long as $N$ is sufficiently large (see Eq.~(\ref{sbar})). 
Although not absolutely necessary, we use the 
Levenberg-Marquardt method \cite{recipes} to determine the numerical
value of the parameter $d_s$ that gives the optimal fit of the Brody
distribution $P_B(S;d_s-1)$ to the histograms. This number $d_f$ can then be
immediately compared to the theoretical value. The purpose of this 
procedure is to test how accurately the Brody distribution
[Eq.~(\ref{Eqn:general})] reproduces the histogram data obtained from 
particular realizations of the model.  
Of course, each realization (in general) yields a unique histogram (and
hence a unique $d_f$),
and so it is more informative to average over several (say $n$) realizations 
and subsequently define $d_f=\bar{d}_f\pm\sigma$, where $\bar{d}_f$ is
the average $d_f$ value obtained from the $n$ realizations 
and $\sigma$ is the standard deviation. 
The percentage error (denoted by $\varepsilon$) of $\bar{d}_f$ 
(obtained from $n=10$ independent realizations) relative 
to $d_s$ is typically under $1\%$. 
We have in fact studied point processes on many of the well-known classical 
fractals in $\mathbbm{R}^2$, $\mathbbm{R}^3$, and $\mathbbm{R}^4$ 
\footnote{Recall that, 
in studies of quantum chaos, the Brody distribution is not meant to be
used beyond the Wigner limit, but in the present context, there
is no such restriction since Eq.~(\ref{Eqn:general}) is
valid beyond the Wigner limit.}, and in almost all cases the accuracy
is comparable. 

The nearest-neighbor energy-level statistics of some quantum systems execute 
a Poisson-to-Wigner transition as the underlying classical dynamics
monotonically change from being completely integrable to completely chaotic.  
One example is the family of Robnik billiards \cite{PR}. 
A monotonic transition between integrabililty and chaos is however quite 
exceptional. More typically, the degree of ``chaoticity'' of  
the classical dynamics (as measured by the volume fraction of phase space 
filled with chaotic trajectories) changes in a complicated way as a system
parameter is varied monotonically, and so the energy-level statistics will 
not undergo a direct transition from Poisson to Wigner. For example,
the energy-level statistics of the hydrogen atom in a van der Waals 
potential undergo a Wigner-Poisson-Brody-Poisson-Brody-Poisson-Wigner
transition as the appropriate system parameter is monotonically varied in a
specified range \cite{HinvanDerWals}. Regardless, in the intermediate
regime between integrability and hard chaos, the 
Brody distribution (albeit a pure surmisal) has often been found to be
a good delineation of the energy-level spacing histogram. 
There are, in fact, systems for which 
the statistical confidence is high (an example is the ripple billiard
\cite{rippL}). This is not to say the Brody distribution 
is now established as a distribution that quantitatively describes  
energy-level statistics in the intermediate regime, but rather that 
after 30 years of pervasive use with considerable success, the Brody
distribution has garnered an undeniable phenomenological significance.
(Of course, other distributions have been 
proposed and used to interpolate between the Poisson and Wigner limits; we 
cite here a few of the more popular distributions \cite{BR,Izzy,LH,CGR,DGOE}. 
These efforts cannot be disregarded, 
but they are not directly relevant to the present discussion.) Given
our present result and the phenomenological status of the Brody distribution 
in studies of quantum chaos, there is a profound implication that transpires: 
\emph{Phenomenologically, 
the energy levels of a typical time-reversal invariant quantum system, whose
dynamics in the classical limit are mixed, have the same nearest-neighbor  
statistics as random points on a fractal with dimension in-between $1$ and 
$2$}. This phenomenological corollary, together with analogous
corollaries in the limiting cases of integrability [Eq.~(\ref{Eqn:Poisson})] 
and hard chaos [Eq.~(\ref{Eqn:Wigner})], offer a new 
phenomenology for quantum chaos: the \emph{nearest-neighbor} energy-level
statistics of a typical time-reversal invariant quantum Hamiltonian
follow the statistics of (i) random 
points on a (one-dimensional) line if the classical limit is integrable; 
(ii) random points on a (two-dimensional) plane if the classical limit
is fully chaotic; and (iii) random points on a fractal set with dimension 
in-between $1$ and $2$ if the classical limit is mixed. 

This phenomenological behavior is quite puzzling. Why should
the energy levels of a quantum Hamiltonian behave (insofar as their 
nearest-neighbor statistics) like random points on a fractal or on a plane? 
This is a very difficult question to answer since there is no direct
connection between point processes and quantum mechanics. Point-process 
models (PPMs) and random-matrix models are both stochastic models, but 
unlike random-matrix models, PPMs do not inherently contain  
any of the structure of quantum mechanics, and so it is difficult to
understand why point-process statistics should have any relation to
energy-level statistics. Surreptitiously, the fundamental link is 
\emph{classical} mechanics. The model
of random points on a fractal can be regarded as a simple stochastic
model for chaotic dynamics on a Poincar\'{e} section. In mixed Hamiltonian
systems, regular and chaotic regions are comingled, and 
the chaotic regions, in particular, are fractal in nature. If we restrict
our scope (at least initially) to two-degree-of-freedom billiard
systems (such as the family of Robnik billiards), then we know that
the chaotic trajectories explore (in a seemingly random fashion) a
fractal subset of the Poincar\'{e} section having dimension in-between 1
and 2. Clearly, the NNSD of the ``chaotic points'' on the section
(corresponding
to a chaotic trajectory) must be Poisson-like in the near-integrable
regime, Wigner-like in the chaotic regime, and some intermediate
distribution in-between Poisson and Wigner in the mixed
regime. The intermediate distribution must also have built in point
repulsion. If random points on fractal sets embedded in
$\mathbbm{R}^2$ really are apt models of Hamiltonian chaos (on a Poincar\'{e}
section), then the intermediate distribution should be the Brody
distribution. If so, then there is an even deeper corollary: 
Phenomenologically, the energy levels of a
typical time-reversal invariant quantum system (whose classical analog is
nonintegrable) have the same nearest-neighbor statistics as the chaotic
trajectories of the underlying classical Hamiltonian. 
Of course, we have not explicitly demonstrated that 
PPMs correctly describe the nearest-neighbor
statistics of chaotic trajectories, and we can only begin to do so 
through numerical experiments. This shall be the
subject of a future paper. The purpose of the above discussion was
merely to introduce the idea of linking PPMs with classical
mechanics, and to outline one of the potential implications. 
For the present, we must settle for the less fundamental, 
but nonetheless far-reaching precursor (italicized above).

\end{document}